\documentclass{emulateapj}
\usepackage{epsfig}
\usepackage{graphicx}
\usepackage{graphics}
\usepackage{color}

\newcommand{\ie}{\emph{i.e.}}

\shorttitle{The formation of double-WD binaries}
\shortauthors{Woods, Ivanova, van der Sluys \& Chaichenets}

\begin{document}

\title{On The Formation of Double White Dwarfs Through Stable Mass Transfer and a Common Envelope}

\author{T.E.\ Woods}
\affil{University of Alberta, Dept.\ of Physics, 11322-89 Ave, Edmonton, AB, T6G 2E7, Canada}

\author{N.\ Ivanova}
\affil{University of Alberta, Dept.\ of Physics, 11322-89 Ave, Edmonton, AB, T6G 2E7, Canada}

\author{M.V.\ van der Sluys\altaffilmark{1}}
\affil{Radboud University Nijmegen, P.O.\ Box 9010, NL-6500 GL Nijmegen, The Netherlands}
\altaffiltext{1}{CITA National Fellow at the University of Alberta}

\author{S.\ Chaichenets}
\affil{University of Alberta, Dept.\ of Mathematical and Statistical Sciences, CAB, Edmonton, AB, T6G 2G1, Canada}

\begin{abstract}
  Although several dozen double white dwarfs (DWDs) have been observed, for many the exact nature of the evolutionary channel(s) by which they form remains uncertain. 
  The canonical explanation calls for the progenitor binary system to undergo two subsequent 
  mass-transfer events, both of which are unstable and lead to a common envelope (CE).
  However, it has been shown that if both CE events obey the standard $\alpha_\mathrm{CE}$-prescription (parametrizing energy loss), 
  it is not possible to reproduce all of the observed systems.
  The $\gamma$-prescription was proposed as an alternative to this description, instead  parametrizing the fraction of angular momentum carried away in dynamical-timescale mass loss.
  In this paper, we analyze \emph{simultaneous} energy and angular-momentum conservation, and show that 
  the $\gamma$-prescription cannot adequately describe a CE event for an arbitrary binary, nor can the first phase of mass loss {\it always} be understood in general as a dynamical-timescale event.
  We consider in detail the first episode of mass transfer in binary systems with initially low companion masses, 
  with a primary mass in the range 1.0--1.3\,$M_\odot$ and an  
  initial mass ratio between the secondary and primary stars of 0.83--0.92. In these systems, 
  the first episode of dramatic mass loss may be stable, non-conservative mass transfer. 
  This strips the donor's envelope and dramatically raises the mass ratio; the considered progenitor binary systems can then evolve into DWDs after passing through a single CE during the second episode of mass loss. We find that such a mechanism 
  reproduces the properties of the observed DWD systems which have an older component with $M\lesssim 0.46\,M_\odot$ and mass ratios between the younger and older WDs of $q\ge1$.

\end{abstract}
\section{Introduction}

A critical phase in the formation and evolution of most close interacting binaries is the so-called common-envelope 
(CE) event, during which the components of a binary system are engulfed by a common gaseous envelope. The resulting drag-like
interaction will then dramatically shrink their orbit \citep{Os76,Pa76}. Depending on the envelope structure and companion masses, 
either the envelope is ejected during this process leaving behind a close binary, or the two stars merge. Dividing the parameter space into 
binaries that survive CEs and those that do not, and determining the final separation for the former, is necessary in order to
calculate the formation rates of low-mass X-ray binaries (LMXBs) and $\gamma$-ray bursts, as well as for LISA and LIGO sources 
\citep{Bel08}. Understanding which evolutionary channels are viable pathways for the creation of double-degenerate systems such as double white dwarfs (DWDs) is one of the
key `sanity checks' available in studying binary evolution, providing tight observational constraints for the otherwise almost entirely theoretical study of the evolution of multiple stars.

The majority of currently observed DWDs consist of relatively low-mass helium dwarfs \citep{SPY01, SPY02, SPY03, SPY04, SPY05}. 
The observed DWDs for which the older companion is less massive than $0.46\,M_\odot$
have an average orbital period of 0.96\,d and an average mass for the (inferred) younger companion of $0.39\,M_\odot$ 
\citep[based  on the double-lined spectroscopic binaries observed prior to this year,][see table 1]{Sluys2006}.

In order to form WD companions of the observed masses, a DWD system must necessarily have undergone two episodes of envelope mass 
loss as each component went through the giant phase.
This can be inferred from the fact that none of these observed WDs are sufficiently massive to have evolved independently to a degenerate state within a Hubble time, and so 
their progenitors, though still low-mass stars, had to be several times more massive than the WDs they produced.
As low-mass giants obey a tight relation between core mass and radius, 
the mass of the younger WD gives the strongest known constraint on the pre-mass-transfer binary configuration.
On the other hand, the initially lower-mass companion must evolve into a white dwarf via a common-envelope process
in the last binary interaction phase, in order to achieve the small separations observed \citep{IL93, Webbink08}. 
However, the physics underlying the evolutionary channel(s) by which DWDs form remains uncertain. 

A variety of explanations have been posited to explain the first stage of mass transfer (MT), in which the initial primary loses its envelope. 
\cite{Han1998} showed that both CE and Roche-lobe overflow (RLOF) are possible in the first episode of the MT. 
Later, it was claimed that this phase cannot be described by stable and conservative RLOF  \citep{Nelemans00, Sluys2006}. 
The suggestion that the first MT episode in the formation history was a CE event also proved problematic: 
when formulated in terms of the binding energy of the envelope, it appeared to imply that the binary must be able 
to eject its envelope with an unphysical (often negative) efficiency \citep{Nelemans00,Nelemans05,Webbink08}. 
Furthermore, the majority of the discovered double-degenerate systems have components of comparable mass, in contradiction 
to the results of past numerical work \citep{Han1998}. 

Recently, a reconciliation was attempted by instead considering angular-momentum balance, re-parametrizing 
the problem in terms of the so-called ``$\gamma$-prescription'' \citep{Nelemans00, Nelemans05}.
In this paper, we demonstrate that the $\gamma$-formalism is not stable against small changes in the binary 
parameters (\S\,3), and that in general it too fails to provide a physical description of an initial, dynamical-timescale phase of evolution which 
is consistent with both energy and angular momentum conservation.

As an alternative, we remove the restriction to dynamical-timescale processes and consider the evolution of red-giant main-sequence (RG-MS) binaries via stable, but {\it non-conservative} 
mass transfer from the primary as the first phase of mass loss (\S\,4,\,5).
In this paper, we limit ourselves to systems where the initial primary has not reached the helium flash. 
We demonstrate that, upon the former secondary reaching RLOF, an ensuing CE phase 
will produce a set of DWD binaries with mass ratios and periods in line with 
the observed DWDs whose older remnant is $\la 0.46\,M_\odot$ (\S\,6). Therefore the stable MT+CE channel provides a natural means for reproducing those DWDs in which the first phase of mass loss appears to have been accompanied by orbital expansion. 

\section{Preliminaries}

We now outline the two canonical prescriptions for treating common-envelope evolution, and develop the standard formalism for 
treating stable mass transfer in the case of Roche-lobe overflow. 

\subsection{Energy balance versus angular-momentum balance}
\label{sec:2.1}

In the standard treatment of CE outcomes,
the final separation of the binary is determined via the ``energy formalism'' \citep{Webbink84}, in which the binding energy of 
the (expelled) envelope is equated to the decrease in the orbital energy $E_\mathrm{orb}$:

\begin{equation}
  E_\mathrm{bind} = E_\mathrm{orb,i} - E_\mathrm{orb,f} 
		= -\frac{ G m_\mathrm{d} m_\mathrm{a}} {2 a_\mathrm{i}} 
			+ \frac{ G m_\mathrm{d,c} m_\mathrm{a}} {2 a_\mathrm{f}}.
\end{equation}

Here $a_\mathrm{i}$ and $ a_\mathrm{f}$ are the initial and final binary separations, 
$m_\mathrm{d}$ and $m_\mathrm{a}$ are the initial star masses (donor and accretor, respectively) 
and $ m_\mathrm{d,c}$ is the final mass of the donor, after losing its envelope.

A parameter $\lambda$ is introduced to characterise the central concentration of the donor envelope:

\begin{equation}
  E_\mathrm{\lambda, bind} = \int_\mathrm{core}^\mathrm{surface} 
				\left (  \frac{Gm}{r(m)} - \varepsilon (m) \right ) dm 
 			= \frac {G m_\mathrm{d} m_\mathrm{d, e}} {\lambda r_\mathrm{d}}.
\label{lambda}
\end{equation}

Here $m_\mathrm{d, e}$ is the mass of the removed (giant's) envelope, $r_\mathrm{d}$ is the radius of the giant star at the onset of the CE
and $\varepsilon$ is the specific internal energy.
$E_\mathrm{bind}$ therefore consists of the potential energy of the envelope and its internal energy,
and can be found directly from the stellar structure for any choice of core mass.
In our calculations, we adopt to include in  $\varepsilon$ only the thermal energy of the gas and the radiation energy,
but not the recombination energy.
There are also alternative definitions for $E_\mathrm{bind}$, where ionization energy \citep[e.g.][]{Han02} or enhanced winds 
\citep[e.g.][]{Soker04} are taken into account. 

Another parameter, $\alpha_\mathrm{CE}$, is introduced as a measure of the efficiency with which energy is transferred from the orbit
into envelope expansion. Invoking energy conservation, one can then find the final orbital separation from

\begin{equation}
  \alpha_\mathrm{CE} {\lambda} 
	\left ( \frac{ G m_\mathrm{d,c} m_\mathrm{a}} {2 a_\mathrm{f}} -\frac{ G m_\mathrm{d} m_\mathrm{a}} {2 a_\mathrm{i}} \right ) 
		=
  \frac {G m_\mathrm{d} m_\mathrm{d,e}} {r_\mathrm{d}}.
\label{allam}
\end{equation}
The obvious bounds on this parameter are $0<\alpha_\mathrm{CE} \leq 1$, since energy cannot be generated. 
The parameter $\lambda$ can be calculated from detailed stellar-structure models. 
For many low-mass giant stars of interest $\lambda\approx 1$, and therefore many authors assume $\alpha \lambda =1$ or $0.5$. 
\citet{vdsluys2010} find $0.027 < \lambda < 1.73$ with a median of $\lambda = 0.86$ for the donors (with masses up to 
$20\,M_\odot$) of a population of 165,000 binaries at the onset of a CE.
Note that this does not hold for massive giants, where $0.004 \lesssim \lambda \lesssim 10$ \citep{Dewi01,podsi03}.

An alternative approach to determine the orbital period after the ejection of the envelope is 
by considering the other available integral of motion, the angular momentum \citep{Nelemans00, Nelemans05}.
This is known as the $\gamma$-formalism and can be expressed as:

\begin{equation}
  \frac{\Delta J_\mathrm{lost}}{J_\mathrm{i}} 
	= \frac{J_\mathrm{i} - J_\mathrm{f}}{J_\mathrm{i}} 
	= \gamma \frac{m_\mathrm{d, e}} {m_\mathrm{d}+m_\mathrm{a}},
\label{gammaform}
\end{equation}

\noindent where $J_\mathrm{i}$ and $J_\mathrm{f}$ are the angular momenta of the initial and the final binaries. Hence, the $\gamma$-parametrization 
describes the specific fraction of the initial angular momentum that is carried away by the envelope as it is lost from the system, 
in terms of a multiplicative factor ($\gamma$) times the fraction of the total mass lost from the system. 

It was shown that the $\gamma$-formalism can reproduce the distributions of the orbital periods and mass ratios in the 
observed DWD systems with a single value of $\gamma=1.5$ for the first mass-transfer event, where the 
 range for possible reconstructed values was found to be between $\sim1.2$ and $\sim2.5$ \citep{Nelemans05}.
At the same time, they showed that no single value of $\alpha_\mathrm{CE}\lambda$ can satisfy all the merger scenarios they considered.
The values that can reconstruct the observed systems scatter from unphysically negative to suspiciously large $\sim 2$, 
with values for $\alpha_\mathrm{CE}\lambda$ skewed to low ($\la 0.5$) values. We note that those low values are rather 
in agreement with expected $\alpha_\mathrm{CE}<1$ and $\lambda$ variation along the giant branch.
The considered systems had primary masses from $1.2$ to $3\,M_\odot$, 
and most of the likely progenitors clustered around $\sim 1.7\,M_\odot$.

\subsection{Orbital evolution during the mass transfer}
\label{sec:2.2}
Assuming a circular orbit, the instantaneous angular momentum about the binary's centre of mass is defined as 
\begin{equation}
\label{eq1}
J = \frac{m_\mathrm{d} m_\mathrm{a}}{\sqrt{m_\mathrm{d} + m_\mathrm{a}}} \sqrt{Ga},
\label{oam}
\end{equation}
where $a$ is the current orbital separation. 

Neglecting the spin angular momentum of the companions, 
a general equation for the orbital evolution of a non-eccentric binary is:
\begin{equation}\label{oam_der}
2 \frac{\dot J}{J} = \frac{\dot a}{a} + 2 \frac{\dot m_\mathrm{d}}{m_\mathrm{d}}+ 2 \frac{\dot m_\mathrm{a}}{m_\mathrm{a}}
-  \frac{\dot m_\mathrm{d}+ \dot m_\mathrm{a}}{ m_\mathrm{d}+m_\mathrm{a}}.
\end{equation}

\noindent Here $\dot J$ and  $\dot a$ are the rates of change of the orbital angular momentum and of the orbital separation, respectively. 
The quantities $\dot m_\mathrm{d}$ and $\dot m_\mathrm{a}$ are the corresponding rates of the total mass loss from the donor and the accretor.

The total mass loss from the donor is $\dot m_\mathrm{d} = \dot m_\mathrm{d,t} + \dot m_\mathrm{d,w}$,
where $\dot m_\mathrm{d,t}$ is the mass transfer rate through $L_1$, and $\dot m_\mathrm{d,w}$
is the mass loss from the donor in the form of a fast isotropic wind. 
As the accretor is less evolved than the donor and is still on the main sequence, its wind mass loss can be neglected.
We introduce the following notations:

\begin{itemize}
 \item $\beta = -{\dot m_\mathrm{a}}/{\dot m_\mathrm{d,t}} $ --- the fraction of the transferred mass that is accreted onto $m_\mathrm{a}$; the conservation factor. With $\beta=1$ the MT is fully conservative.

 \item $\delta = {\dot m_\mathrm{d, w}}/{\dot m_\mathrm{d}}$ ---  the fraction of the total donor mass that was lost in the form of a fast isotropic wind. 
\end{itemize}

With these notations, the total mass loss from the system is
\begin{equation}\label{tot_massloss}
  \dot m_\mathrm{d}+ \dot m_\mathrm{a} =  (1-\beta+\beta\delta)\, \dot m_\mathrm{d},
\end{equation}
and the mass accreted is
\begin{equation}\label{acc_mass}
  \dot m_\mathrm{a} =-\beta  (1-\delta)\, \dot m_\mathrm{d}.
\end{equation}

The specific angular momenta of the two binary members, and hence of the mass that is carried away from them 
(from the accretor star as transferred material that is isotropically re-emitted, and from the donor star in the form of a fast isotropic wind), are:
\begin{equation}
  h_\mathrm{a} = \frac{m_\mathrm{d}}{m_\mathrm{a}} \frac{J}{m_\mathrm{d}+m_\mathrm{a}},    \hskip1cm              h_\mathrm{d} = \frac{m_\mathrm{a}}{m_\mathrm{d}} \frac{J}{m_\mathrm{d}+m_\mathrm{a}},
\end{equation}
and the total loss of angular momentum from the binary system can be written as:
\begin{equation}
  \label{amloss}
  \frac{\dot J}{J} = \frac{\dot m_\mathrm{d} }{m_\mathrm{d}+m_\mathrm{a}} \left [\frac{m_\mathrm{d}}{m_\mathrm{a}} (1 - \delta) (1-\beta)  +  \frac{m_\mathrm{a}}{m_\mathrm{d}}\delta  \right ].
\end{equation}

Combing eq.(\ref{oam_der}) and eq.(\ref{amloss}), we find that the orbital evolution for our calculations of evolutionary sequences during RLOF is:
\begin{equation}\label{orbit}
\frac{\dot a}{a} = \frac{\dot m_\mathrm{d} }{m_\mathrm{d}} \,
\frac{(1-\delta)(2m_\mathrm{d}^2 - 2m_\mathrm{a}^2 +\beta m_\mathrm{d}m_\mathrm{a}) - m_\mathrm{d} m_\mathrm{a}}{m_\mathrm{a} (m_\mathrm{d}+m_\mathrm{a})}.
\end{equation}
Note that in our calculations we only choose to fix the constant conservation factor $\beta$ for each individual binary evolution, but not $\delta$, 
which is always a function of the current evolutionary state and MT rate. 
During the initial MT phase, when MT often operates on a timescale close to thermal, 
and the MT rate is high, $\delta$ is effectively 0. 
The inclusion of $\delta$ becomes more important during the late stage of the MT, when the MT operates on the nuclear timescale  
(for more details, see \S\,5).

\section{Analysis of the $\gamma$-formalism}
\label{sec:three}

\cite{Nelemans00} analyzed the formation of DWD by attempting to reconstruct known systems. 
The authors found that for the second phase of mass loss, a CE event with $\alpha_\mathrm{CE} \lambda\simeq 2$ 
could be fit to the majority of observed systems. However, one should assess the implications of this with caution 
as we would expect use of a single-valued parameter to be inappropriate in the general case, as there is little 
reason to assume that such a parameter should not vary under differing conditions.  Specifically, in the case of 
the $\alpha_\mathrm{CE} \lambda$ formalism, $\lambda$ is known to vary along the giant branch, and it would seem 
a somewhat awkward contrivance to suggest a single envelope removal efficiency is shared among all binaries undergoing 
a CE. Such considerations are essential in order to provide more substantive results in population synthesis. 
\cite{Nelemans00} find plausible values for $\alpha_\mathrm{CE} \lambda$ ranging from 0.5 to 3 for different observed 
cases. We take a value of $\alpha_\mathrm{CE}$ to be reasonable so long as it is in the physically meaningful range 
between $0$ and $1$.

The same reconstruction analysis in \cite{Nelemans00} has also shown that the first phase of mass loss 
can be explained with a CE only if $-15 \leq \alpha_\mathrm{CE}\lambda \leq -5$, 
well outside any physically acceptable range. 
However, this assumes that the only relevant source of energy is the binding energy of the envelope. 
For the orbit to widen in the course of mass loss, Eq.(\ref{allam}) requires an influx of energy outside of that which 
is available on a dynamical timescale.
This is not at all unphysical for dynamically stable MT via RLOF, which may take place on the thermal or nuclear timescale 
and is driven by the nuclear burning of the donor. Therefore the cause for this unphysical behaviour would seem to be 
immediately clear; stable MT via RLOF can in many cases provide a clear solution for the removal of the initial primary's 
envelope. 

However, the possibility that stable conservative MT could account for this stage was previously ruled out. 
\cite{Nelemans00} empirically analyzed Algol type MT, specifically MT initiated while the donor is still on the main sequence.
In a later study \citep{Sluys2006}, conservative MT was considered and rejected once more.
We note however that neither of the studies performed the detailed evolution for giant donors at the start of the MT.
\cite{Nelemans00} empirically analyzed only stars with less than 50\% of the outer envelope being convective at RLOF
(e.g., for a $1.2\,M_\odot$ star, this limit means that the star is as small as $\sim 2.6\,R_\odot$  and has barely left the
bottom of the giant branch). In \cite{Sluys2006}, it was adopted that all stars beyond the bottom of the giant branch will start 
unstable MT (see their condition for instability 7), so the detailed MT calculations were performed only for stars
that have not become giants yet upon reaching RLOF. We note that in both studies the donors could become giants after the start of the MT
and continue MT stably, but the initial conditions limited the donors at the end of the MT to rather early giants. 

An alternative formulation, in which the envelope of the initial 
primary is lost without significant spiral-in, was formulated in terms 
of a new parameter $\gamma$ (see \S \ref{sec:2.1}).  
It was found that $\gamma=1.5$ allows for widening or very mild shrinkage 
in the first CE phase (for mass ratios very close to unity) but extreme shrinking in the second.
As a result, it has been claimed in \cite{Nelemans05} that this
single value of $\gamma$  can explain {\it all} observations and therefore can be used
to predict the outcome of a CE, as well as any other dynamical-timescale evolution, for any arbitrary binary population, 
including massive binaries. 

The $\gamma$ formalism can indeed lead to orbital expansion, as well as an increase in the total energy, for mass ratios close to 
$1$. However, if interpreted as a dynamical-timescale process, this increase must prove just as unphysical as a negative 
efficiency in the $\alpha_\mathrm{CE} \lambda$ formalism. 
In the following section we will also demonstrate that a single-valued $\gamma$ is not robust against small perturbations 
in orbital parameters, and is thus unable to explain both evolutionary stages in all instances. Therefore, when attempting 
to describe both phases of DWD formation simultaneously, the $\gamma$ formalism is subject to similar fundamental physical 
problems as those found using the $\alpha_\mathrm{CE} \lambda$ prescription. Rather, in $\S$ \ref{1stMassLoss} we demonstrate that 
stable but non-conservative mass transfer can account for the expansion in the initial phase, bypassing the need for any 
parametrization of this stage. 

In order to understand what the $\gamma$ mechanism implies, and to contrast it with the $\alpha_\mathrm{CE} \lambda$ 
formalism in a double-CE scenario,
let us analyze the energy and angular-momentum balance during a first and second CE of a binary 
system that is a potential progenitor of a DWD system.
After the first CE, this system will presumably evolve to a 
second (and last) CE event, according to the double-CE model via the $\gamma$ mechanism
by \citet{Nelemans00, Nelemans05}.  In their scenario, at the start of the second CE event
a giant donor has a mass of about $1.7-2.2\,M_\odot$ and
the original primary star has already formed a WD with a mass of $\sim 0.35-0.55\,M_\odot$.

\subsection{Common envelope via $\alpha$-formalism}
\label{sec:example1}

\begin{figure}
  \includegraphics[height=.36\textheight]{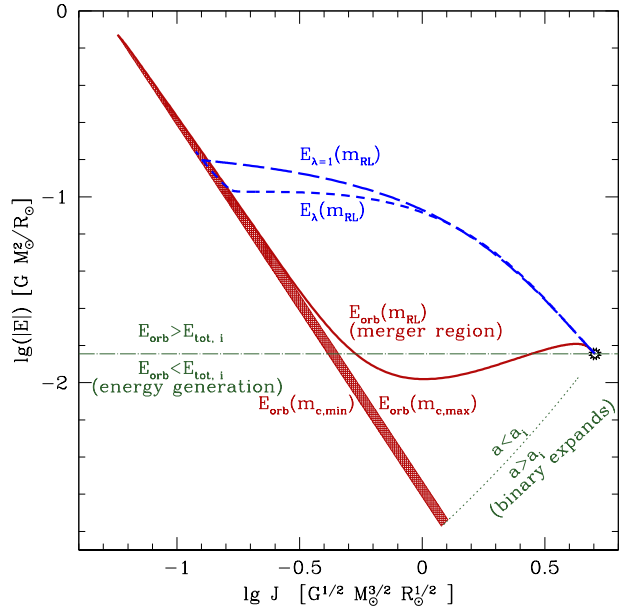}
  \caption{
    Orbital angular momentum $J$ and energy $E$ for a CE in a $2\,M_\odot + 0.5\,M_\odot$ binary
    ($r_\mathrm{d}=23\,R_\odot$ and $m_\mathrm{d,X}=0.337\,M_\odot$).
    The black star indicates the state of the binary at the onset of the CE.
    The solid line (red) shows the Keplerian $E_\mathrm{orb}- J_\mathrm{orb}$ relation for the final binary
    assuming that the mass of post-CE remnant consists of all the RG mass that was originally 
    contained within the final Roche lobe ($m_\mathrm{RL}$), in other words, the maximum possible
    remnant mass for this orbit.
    The shaded region contains Keplerian orbits for core masses
    bounded by $m_\mathrm{c,min} < m_\mathrm{d, c} < m_\mathrm{c, max}$ (see the text).  
    A final state of the  post-CE binary must lie within this area. 
    The dashed lines (blue) trace the minimum energy expenditures to release the envelope
    with, for $\lambda$ calculated using the eq.\,(\ref{lambda}) and for $\lambda=1$.
    Dashed-dotted green line separates regions where the post-CE binary has more or less energy than at the start of the CE.
    Dotted green  line separates regions where the post-CE is wider ($a>a_\mathrm{i}$) than at the start of the CE and where it shrunk ($a<a_\mathrm{i}$).
    \label{alpha}
  }
\end{figure}

As an example, we consider a binary with a giant of $2\,M_\odot$ and a WD of $0.5\,M_\odot$,
where the CE is initiated when the giant has a radius of $21\,R_\odot$.
The states of the binary system in the orbital-energy--angular-momentum ($E-J$) plane 
are plotted in Fig.\,\ref{alpha}.
The location of the initial state is marked with a star.
For this state the orbital energy $E_\mathrm{orb, i}$ and the rotational energy of the giant's envelope
(assumed to be synchronized with the initial orbit), 
as well as the orbital angular momentum $J_\mathrm{orb, i}$ and angular momentum of
the giant's envelope are taken into account. 

The thick solid line is determined by repeatedly removing the outermost mesh point from the structure model,
assuming that the radius of the resulting star is equal to the radius coordinate the new outer mesh point had in
the original model (\ie, the model doesn't expand), equating this radius to the Roche-lobe radius 
and then computing the orbital properties of the resulting binary. 
We also note that in this figure, the adiabatic expansion of the post-CE remnant due to mass loss was not taken into account
\citep[see discussion in][]{Deloye10,Ivanova10}, as such, the solid line provides the closest possible orbits
and, realistically, the position of the binary can only be below this line.
The region where the energy is below its initial value is an impossible final configuration for a post-CE binary, 
and is present in the figure only as a limiting case, as such a configuration is not physically possible.

For any given core mass, Kepler's law constrains the orbits to a straight line parallel 
to the shaded region, which is bounded by the minimum and maximum core masses.
Our choice of the \textit{lower} bound on the possible core mass $m_\mathrm{c, min}$
is the hydrogen-exhausted core $m_\mathrm{d, X}$ (the region where $X < 10^{-10}$) and is rather standard.
For the \textit{upper} bound on the possible core mass $m_\mathrm{c, max}$ we choose the minimum between
the central mass which contains less than 10\% hydrogen \citep[as in][]{Dewi00} and
the bottom of the outer convective envelope ($m_\mathrm{BCE}$). 
The definition of the post-CE core is an important problem in itself.
We should stress that there is no strong constraint from either observations or theory 
which dictates that the post-CE remnant immediately after the dynamical event will have 
the exact same mass as the WD companion that is observed much later.
As such, the mass of a post-CE remnant could be anywhere
between our minimum and maximum core-mass definitions. A stronger constraint
on the post-CE core mass is proposed in \cite{IvaCha09,Ivanova10}.

The solid line in Fig.\,\ref{alpha} represents the \textit{closest possible orbit} 
from Roche-radius considerations: 
only the states below the solid line are possible, and the actual state depends on the degree of the post-CE detachment.
Only the strip bounded from above by the thick solid line and lying within 
the shaded region is permitted for a self-consistent final post-CE binary.
We note from the Fig~\ref{alpha} that, depending on the choice of the post-CE remnant mass, 
the minimum possible semi-major axis of the post-CE binary can vary by a factor of 30.

The post-CE binary can not have more energy than the initial binary and still satisfy conservation of energy (see dashed-dotted green line on Fig.~\ref{alpha}), since some energy must be used to expel the common envelope
to infinity.
We can also find the position that a final binary would have if it kept the same orbital separation, for any valid post-CE mass
of the donor $m_\mathrm{c, min}<m_\mathrm{d,c}<m_\mathrm{d}$. Clearly, a widened orbit would violate energy conservation.

We also plot the energy ``expense'', the initial energy value
plus the energy needed to disperse the envelope above a given core mass, as the dashed lines in Fig.\,\ref{alpha}  
(assuming $\alpha=1$ --- the most efficient case possible, a smaller value would shift the curve higher).
The different points on these tracks are determined by removing the outer mesh point,
computing the amount of energy needed to expel that mass shell to infinity, and removing
that energy from the binary orbit. 
The angular momentum for these states is computed using the remaining masses and orbital energy
to satisfy Kepler's law.
As this energy expense is the smallest possible, the final binary state should
lie anywhere \textit{above} the dashed lines for a given core mass.
For comparison, we show a solution where $\lambda$ is calculated from the stellar structure, 
and where $\lambda=1$ is assumed. 

Energy considerations limit the range of possible post-CE binary separations,
as there is a minimum amount of energy that needs to be extracted.
Nonetheless, for a properly computed $\lambda$, final orbital separations can vary 
over a factor of 10, depending on the choice of core mass.
The whole range of possible states is now bounded by the blue dashed line to the bottom,
the solid red line to the top and the shaded red area.
When the CE is initialised at a different giant radius (\textit{i.e.}, at a different orbital period), 
and accordingly a different $m_\mathrm{d,X}$, the picture is qualitatively similar, although the 
uncertainty that is introduced by the core-mass definition could vary.

\subsection{Common envelope via $\gamma$-formalism }

\label{sec:example2}

We now construct an example that fits the $\gamma$-parametrisation introduced
in \citet{Nelemans00} and further studied in \citet{Nelemans05},
and show that it is not robust against small changes in binary parameters.
The latter study determined the range of values for $\gamma$ to be between $\sim 1.2$ and up to $4$
in the first episode of the MT and between  $\sim 0.5$ and $3$ in the last episode of the MT, 
with the most likely value of $\gamma\approx1.5$ --- \textit{i.e.}, every gram of the 
escaped envelope material carries away 1.5 times the average specific angular momentum 
of the binary.

\begin{figure}
  \includegraphics[height=.36\textheight]{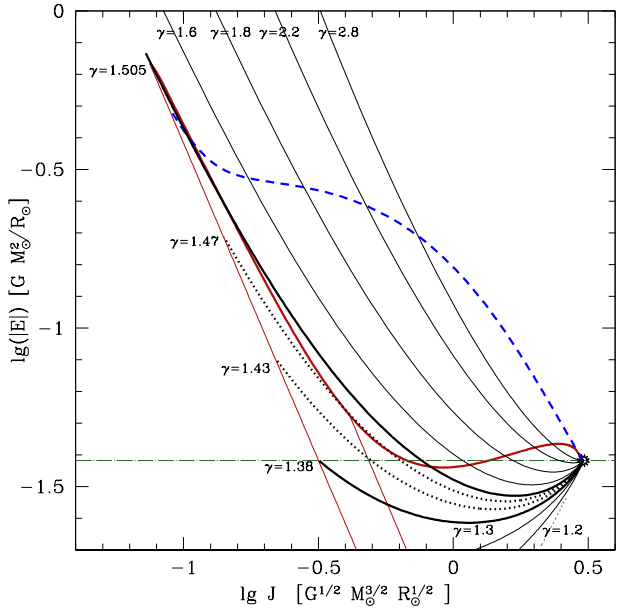}
  \caption{
    Orbital angular momentum and energy for a $2\,M_\odot + 0.5\,M_\odot$ binary
    ($r_\mathrm{d}=86\,R_\odot$ and $m_\mathrm{d,X}=0.38\,M_\odot$).
    Thick and thin solid red lines show the only possible final binary configurations for various
    adopted core masses, for the comparison we also show the the minimum energy expenditures to release the envelope 
    (blue dashed line, see also the caption of Fig.\,\ref{alpha}).
    Black solid and dotted lines indicate possible final binary configurations, assuming angular momentum lost
    in accordance with the $\gamma$-formalism, where thick black solid lines show minimum and maximum $\gamma$ 
    that make a physically viable binary, and dotted black lines show values of $\gamma$ that lead to formation
    of physical viable binaries for all range of the core masses. Other lines' styles are as in Fig.\,\ref{alpha}.
    \label{gamma1}
  }
\end{figure}

Let us choose a $2\,M_\odot$ giant, evolved to a hydrogen-exhausted core mass $m_\mathrm{d, X}\sim 0.4\,M_\odot$,
a common case in the study of \citet{Nelemans05}, where many double WD binaries
have an older companion of $0.5\,M_\odot$ and a younger WD of $0.4\,M_\odot$.
In Fig.\,\ref{gamma1} we again plot the final binary configurations in the $E-J$ plane, in accordance 
with the $\gamma$-formalism. 
The black tracks are constructed by subsequently removing mass-mesh points from the
surface of a stellar-structure model.  The value for $J$ is computed from the original angular momentum
of the binary, and by applying the $\gamma$-formalism.  The value for $E$ is computed then from Kepler's law.
The only self-consistent solutions are those that fall within the strip delineated by the two thin solid red lines (corresponding to the
red shaded area in Fig.\,\ref{alpha} and described in \S~\ref{sec:example1}) for the 
adopted core-mass range.  The solutions also are bound from above by the solid red line, where the 
binary is bound to merge, and by the dashed-dotted green line, which denotes the initial binary energy: below this line, the $\gamma$-formalism essentially predicts energy generation.

From Fig.\,\ref{gamma1}, we see that the track for $\gamma=1.5$ roughly coincides with a possible 
final binary configuration for these particular companion masses. \textit{By coincidence}, the $\gamma\approx 1.5$ 
solution crosses the final binary configuration at approximately the location mandated by energy consideration.
However, we will show below that this is no longer the case for a giant of the same mass but with a different
core mass.  The self-consistent solutions in this example are those with $\gamma \approx 1.43-1.47$ (dotted black lines),
where for some core masses values between 1.38 and 1.505 could also give physically valid solutions (thick solid black lines).
Note that the change in $\gamma$ by $\pm0.05$ around $\gamma=1.45$, for the same core mass, results in final orbits being different by a factor of 10,
while a bigger change in $\gamma$ leads to either merger, or to an unphysical energy generation.
As such, it is not possible for one system to have a large range of $\gamma$ giving physically valid solutions, 
such as from 1.2 to 3, as in \cite{Nelemans05}.

As we indicated, the consistency of the solution considered above is coincidental.
We will now consider two examples where a CE with $\gamma \approx 1.5$ does not lead to a
self-consistent solution.

\begin{figure}
  \includegraphics[height=.36\textheight]{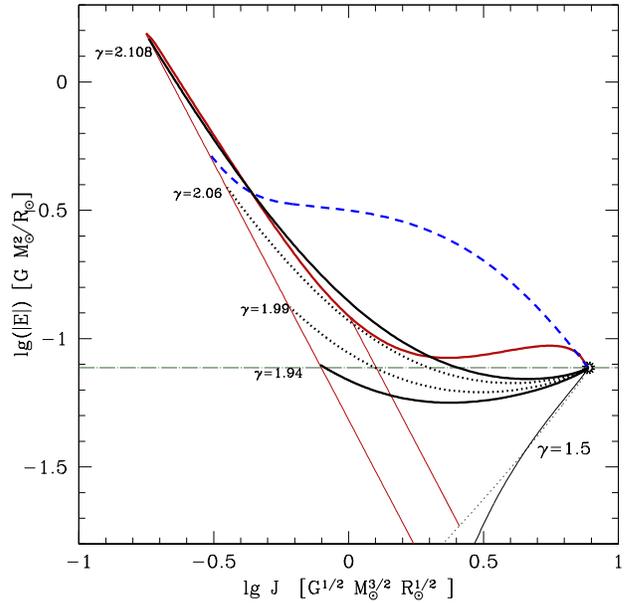}
  \caption{Orbital angular momentum and energy for a $2\,M_\odot + 1.5\,M_\odot$ binary
    ($r_\mathrm{d}=86\,R_\odot$ and $m_\mathrm{d,X}=0.38\,M_\odot$).
    Line styles are as in Fig.\,\ref{gamma1}.
    \label{gamma2}
  }
\end{figure}

First, we take the same $2\,M_\odot$ giant with the same $m_\mathrm{c, min}\sim 0.4\,M_\odot$,
but with a $1.5\,M_\odot$ companion (Fig.\,\ref{gamma2}), similar to some initial binaries considered to be progenitors for DWDs in \cite{Sluys2006},
and consider how the $\gamma$ formalism treats the first phase of MT here.
This binary would have a self-consistent  post-CE binary separation only for $\gamma\approx1.99-2.06$ (close to values of $\gamma$ used
in some similar systems in \cite{Sluys2006}, see their Table 6). 
With $\gamma=1.5$, this binary would extract too little angular momentum to remain an energy-consistent binary:
the binary is \textit{even less bound} than it was at the beginning --- the same effect as having negative $\alpha$ or a stellar wind;
apparently energy is generated during the mass removal to infinity. As a consequence, the binary is becomes wider during the mass loss.
Speaking physically, it means that, if $\gamma=1.5$ can reconstruct some systems, in these systems the mass loss process was not 
conserving energy, and as such could not proceed on a dynamical-timescale and be called a common envelope.
This is a clear indication of a stable mass transfer event and the MT phase has to be studied and understood using the appropriate tools.

\begin{figure}
  \includegraphics[height=.36\textheight]{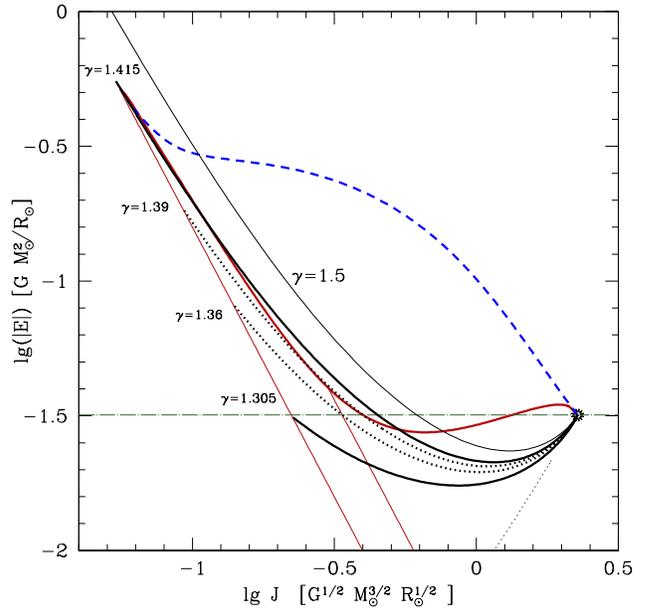}
  \caption{Orbital angular momentum and energy for a $2\,M_\odot + 0.35\,M_\odot$ binary
    ($r_\mathrm{d}=86\,R_\odot$ and $m_\mathrm{d,X}=0.38\,M_\odot$).
    Line styles are as in Fig.~\ref{gamma1}.
    \label{gamma3}
  }
\end{figure}

Another example involves a binary with a smaller companion ($0.35\,M_\odot$, Fig.\,\ref{gamma3}).  Keplerian 
solutions can be found for $\gamma \approx 1.36-1.39$, where the largest acceptable solution is
$\gamma\approx 1.415$.  With $\gamma=1.5$, this binary has to extract so much angular momentum 
that the post-CE binary clearly has to merge if the realistic core sizes are taken into account, 
whereas energy consideration  indicates that a merger is not necessary, though it is at a threshold.

We conclude that the $\gamma$-parametrisation with a unique value of $\gamma$ can not be used for arbitrary companion masses.
A small variation in the value of $\gamma$ covers the whole range of possible solutions --- from unphysically expanded,
to all possible physically acceptable configurations, and to mergers.
As such, one {\it cannot} employ the $\gamma$-formalism using $\gamma = 1.5$ for an arbitrary binary, as $\gamma$ values must be fit for each system.
To resolve the apparently negative $\alpha_\mathrm{CE}$ values required to explain the formation 
of many DWD systems, we propose that instead of a first common-envelope event, stable mass transfer 
takes place. 
While it has been claimed \citep{Nelemans00, Sluys2006} that an initial phase of fully conservative MT 
is incompatible with the observed distribution of DWDs, this was demonstrated only for donors that started mass transfer near the beginning of the RGB. 
The degree of orbital expansion is greatly reduced if one introduces an inefficiency in the process --- non-conservative MT.
In this case, the second CE can proceed with physically reasonable values for $\alpha_\mathrm{CE}$.
To avoid an initial CE, the first episode of mass transfer must proceed in a dynamically stable manner.

\section{Stability of mass transfer}

Whether, once started, MT will proceed in a stable or unstable manner depends on 
the response of the donor and its Roche lobe to the MT.
Mass loss is a perturbation that brings a star out of hydrostatic and thermal equilibrium.
In order to re-establish these equilibria, the star will either expand or contract,
first restoring hydrostatic equilibrium, and then, on a longer timescale, thermal equilibrium.
Like the star's radius, its Roche-lobe radius also changes in response to the MT.

As a stability condition, we thus demand that upon MT the new donor radius
should remain within the donor's new Roche lobe.
The \textit{dynamical} stability of the MT depends on the \textit{adiabatic} response of the stellar radius to mass loss. 
To consider stability in detail, we will use the linear stability analysis following \citet{HW87, SPV97}. 
We define the adiabatic mass--radius exponent (the donor's response to the mass loss
on an adiabatic timescale) as
\begin{equation}
  \zeta_\mathrm{ad}  \equiv \left ( \frac{d\log\,r_\mathrm{d}}{d\log\,m_\mathrm{d}} \right )_\mathrm{ad},
\end{equation}
and the Roche lobe's mass--radius exponent (the Roche-lobe response to the MT) as
\begin{equation}
  \zeta_\mathrm{RL}  \equiv  \frac{d\log\,r_\mathrm{RL}}{d\log\,m_\mathrm{d}}.
\end{equation}
The criterion for dynamical stability of MT then can be written as  $\zeta_\mathrm{ad} \ge \zeta_\mathrm{RL}$.

If this criterion is satisfied, the donor is able to recover its hydrodynamical equilibrium 
while remaining within its Roche lobe.
It will then try to recover its thermal equilibrium on the Kelvin-Helmholtz timescale. 
The change in its equilibrium radius
can be expressed with the use of the equilibrium mass--radius exponent
\begin{equation}
\zeta_\mathrm{eq}  \equiv \left ( \frac{d\log\,r_\mathrm{d}}{d\log\,m_\mathrm{d}} \right )_\mathrm{eq}.
\end{equation}
If, in addition to the dynamical-stability condition, the condition $\zeta_\mathrm{eq} \ge \zeta_\mathrm{RL}$ 
also holds, then MT will proceed on the nuclear-evolution timescale in a secularly stable manner. 
If $\zeta_\mathrm{ad} \ge \zeta_\mathrm{RL} > \zeta_\mathrm{eq}$,
MT will be driven by thermal readjustment and will proceed stably on the 
thermal timescale ($\tau_\mathrm{th}$).
In the case of a red giant, the radius of the star in complete equilibrium is a function
predominantly of its core mass, and as such its thermal response is usually adopted 
to be $\zeta_\mathrm{eq} \approx 0$, though it can vary.
We will now consider these responses in more detail.

\subsection{Adiabatic response}

The adiabatic mass--radius exponent for giants can be obtained considering the case 
of a condensed polytrope with $n=3/2$ \citep{HW87}.  The value of 
$\zeta_\mathrm{ad}$ can then be found to within 1\% accuracy using the formula from \citet{SPV97}:
\begin{equation}\label{ad_resp}
  \zeta_\mathrm{ad} = \frac{2}{3}\frac{M_\mathrm{c}}{1-M_\mathrm{c}} - \frac{1}{3}\frac{1-M_\mathrm{c}}{1+2M_\mathrm{c}} - 0.03M_\mathrm{c} + 0.2\frac{M_\mathrm{c}}{1 + (1 - M_\mathrm{c})^{-6}},
\end{equation}
where $M_\mathrm{c} \equiv m_\mathrm{c}(M_\odot)/m$ is the mass fraction of the helium core. 

\subsection{Roche-lobe response}
\label{sec:rl_response}

We expand the logarithmic derivative of the orbital separation with respect to the donor mass as:
\begin{equation}
  \zeta_\mathrm{RL} =  \frac{\partial \ln a}{\partial \ln m_\mathrm{d}} + \frac{\partial \ln (r_\mathrm{RL}/a)}{\partial \ln q}\frac{\partial \ln q}{\partial \ln m_\mathrm{d}}.
  \label{eq:zeta_rl}
\end{equation}
Here $q=m_\mathrm{d}/m_\mathrm{a}$ is the mass ratio.
The first term, $\partial \ln a/ \partial \ln m_\mathrm{}$, is solely due to the mass loss or MT and can be found using eq.(\ref{orbit}):
\begin{equation}
  \frac{\partial \ln a}{\partial \ln m_\mathrm{d}} =  \frac{m_\mathrm{a}}{\dot m_\mathrm{d}} \frac{\dot a}{a}.
\end{equation}
We will be most concerned with stability at the onset of a fast initial MT phase (see $\S 4.3$, $\S5.3$). In this case we may assume that the MT rate from the 
donor exceeds the wind mass-loss rate significantly, so that for the purpose of this stability 
analysis we can assume that $\delta=0$ (See \S~\ref{sec:2.2}).
Then:
\begin{equation}
  \frac{\partial \ln a}{\partial \ln m_\mathrm{d}} =  \frac{ 2m_\mathrm{d}^2 - 2m_\mathrm{a}^2  - m_\mathrm{d}m_\mathrm{a} (1-\beta)}{m_\mathrm{a} (m_\mathrm{d}+m_\mathrm{a})},
  \label{eq:dlna_dlnm}
\end{equation}
where $\beta$ is the fraction of the transferred mass that is accreted by the secondary star.
We note that the presence of wind makes MT more stable, as the orbit always expands 
in response to wind loss. As such, with $\delta=0$ we would obtain a stricter criterion for 
MT stability. The above equation demonstrates that the Roche-lobe response 
is a function of the MT conservation factor; in general, $\zeta_\mathrm{RL} \approx \zeta_\mathrm{RL}(\beta \mathrm{,q})$.

The second term in Eq.\,\ref{eq:zeta_rl} consists of the Roche lobe's response to the change in
mass ratio, which can be described using Eggleton's approximation \citep{Egg83}:
\begin{equation}
  \frac{\partial \ln (r_\mathrm{RL}/a)}{\partial \ln q} = \frac{2}{3} - \frac{q^{1/3}}{3}\frac{1.2q^{1/3} + 1/(1 + q^{1/3}) }{0.6q^{2/3} + \ln (1 + q^{1/3})}
  \label{eq:dlnq_dlnm}
\end{equation}
\citep[see also][]{SPV97}, and the response of the mass ratio to the change in donor mass
\begin{equation}
  \frac{\partial \ln q}{\partial \ln m_\mathrm{d}} = 1 + \beta \, \frac{m_\mathrm{d}}{m_\mathrm{a}}.
  \label{eq:dlnrl_dlnq}
\end{equation}

We can then compute $\zeta_\mathrm{RL}$ from Eq.\,\ref{eq:zeta_rl}, using the necessary parts from
Eqs.\,\ref{eq:dlna_dlnm}, \ref{eq:dlnq_dlnm} and \ref{eq:dlnrl_dlnq}.

\subsection{Stability boundary}
\label{sec:stability}

\begin{figure}
\label{unrealistic}
  \includegraphics[height=.35\textheight]{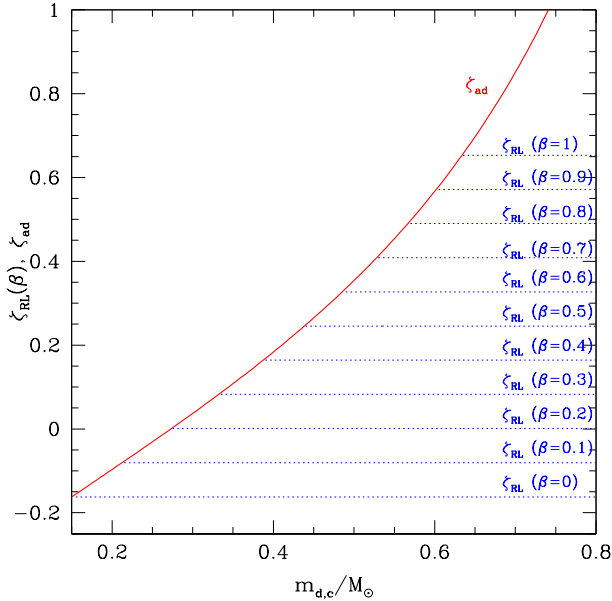}
  \caption{ Adiabatic and Roche-lobe mass--radius exponents for a red giant of $1.2\,M_\odot$ with a companion of $1.1\,M_\odot$,
    as a function of the red-giant core mass. The primary mass is held constant during its evolution up to RLOF. Roche-lobe mass--radius exponents are shown for different cases of mass conservation.
    \label{zetas}
  }
\end{figure}
\begin{figure}

  \includegraphics[height=.35\textheight]{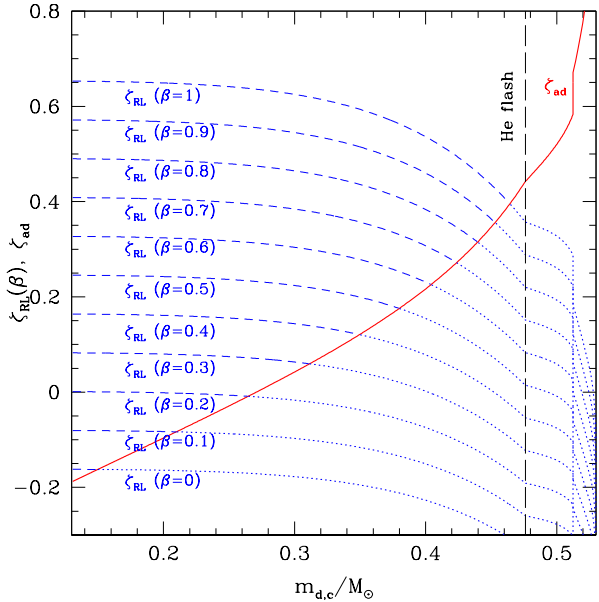}
  \caption{ Adiabatic and Roche-lobe mass--radius exponents for a red giant of $1.2\,M_\odot$ with a companion of $1.1\,M_\odot$,
    as a function of the red-giant core mass. The primary has a metallicity of Z=0.03 and is evolved with wind mass loss. 
    Roche-lobe mass--radius exponents are shown for different cases of mass conservation.\label{realistic}
  }
\end{figure}

By comparing $\zeta_\mathrm{ad}$ and $\zeta_\mathrm{RL}(\beta)$ we can find a $\beta_\mathrm{max}(q,m_\mathrm{d,c})$, such that
for all $\beta\le \beta_\mathrm{max}$, $\zeta_\mathrm{ad}\ga \zeta_\mathrm{RL}(\beta)$ and 
therefore MT will be dynamically stable. We visualize this in Fig.\,\ref{zetas} 
for the case of a $1.2\,M_\odot$ red giant, with unchanging mass during its evolution, and a $1.1\,M_\odot$ companion at the start of MT, 
demonstrating how this condition changes for increasingly evolved red giants.

It can be seen from Fig.\,\ref{zetas} that in all the cases of fully non-conservative MT
 ($\beta=0$), the MT is dynamically stable. 
We also note that fully conservative MT is dynamically stable for cores with masses $m_\mathrm{d,c} \ga 0.63\,M_\odot$,
and in this case MT will proceed on the thermal timescale.

In Fig.\,\ref{realistic} we see that in the more realistic case in which we account for mass loss (including wind loss), as well as z = 0.03, our range of stability is extended even further, to the point of allowing fully conservative MT for large core masses ($M_\mathrm{d,c} \ga 0.46\,M_\odot$). With decreasing 
mass RLOF-driven MT is increasingly stable, e.g.\ for a $1.1+1.0\,M_\odot$ RG-MS system conservative MT is stable for
$M_\mathrm{c} \ga 0.42\,M_\odot$.

Of primary interest to us at the present moment are red giants that are potential progenitors of the inferred older companions of observed DWD systems,
hence our focus on systems in which the donor core mass reaches $\approx 0.35-0.46\,M_\odot$ (before the He flash). 
Note that RLOF-driven MT may often run on the nuclear timescale, 
such that the core mass can grow significantly during this phase.  Hence
we must consider those systems in which MT begins with a core mass as low as $0.25\,M_\odot$ and up (see {\S 5.3.2}).  

Using the above condition for stable MT, we can define an upper boundary on the mass ratio for any given core mass of the donor in 
order for the initial MT phase to proceed in a stable manner. This upper boundary varies with conservation 
factor through $\zeta_\mathrm{RL}$'s dependence on $\beta$ (see Eq.\,\ref{eq:zeta_rl}). For the range of core masses 
of interest here, we find  
that if MT occurs with a conservation factor no greater than 
$\beta\approx0.3-0.5$ (depending on initial mass ratio), it will be dynamically stable.

We also note that the stability criterion we use may not be final, as it is based on values of $\zeta_\mathrm{ad}$ for condensed polytropes, not for real giants. 
As far as we know this is a more restrictive criterion than the ones typically used in the literature, which are frequently based on the mass ratio.
For example, for giant donors $q=1.2$ is considered to be a threshold in \citet[][see the references therein]{Bel08}.
This condition is derived from detailed MT calculations resulting in runaway rates. 
Adopting the criterion based on the comparison of 
$\zeta_\mathrm{ad}\ga \zeta_\mathrm{RL}(\beta)$, we are at the  most conservative limit, and even this limit predicts that some 
systems can have stable, fully conservative MT. 
We also check when the MT rate grows exponentially (indicating instability) in our MT sequences and compare it to the 
predicted value of $\beta_\mathrm{max}$.

\subsection{Thermal equilibrium response and the end of thermal-timescale mass transfer}

In the case of a red giant, the radius of the star in complete
equilibrium is a function predominantly of its core
mass, and its thermal response is usually considered to be $\zeta_\mathrm{eq} \approx 0$.
However, the detailed comparison of giants of different total masses,
but with identical core masses, indicates a dependence on the total mass as well \citep[see e.g.\ Fig.\,1 in][]{Sluys2006}.
In particular, for giants with masses $\la 1.2\,M_\odot$,
$\zeta_\mathrm{eq}$ is negative and a function of the giant's core mass. 
To find it more precisely, one can build a sequence of stellar models in thermal equilibrium with the same chemical composition and the core mass,
starting with $1.2\,M_\odot$ and then decreasing the mass. 
This is done by imposing a very slow mass loss on a $1.2\,M_\odot$ giant and switching off the chemical evolution 
between the relaxed stellar models.

We find that a giant with a $0.37\,M_\odot$ core has $\zeta_\mathrm{eq}\approx -0.4$ as long as the giant mass is between $0.6\,M_\odot$ and $1.2\,M_\odot$.
For masses below $0.6\,M_\odot$, $\zeta_\mathrm{eq}$ becomes positive.
We know that as long as  $\zeta_\mathrm{RL}> \zeta_\mathrm{eq} $, MT proceeds on the thermal timescale,
and it will switch to nuclear-timescale MT when $\zeta_\mathrm{RL} < \zeta_\mathrm{eq}$.
We can then find the mass ratio $q_\mathrm{crit}$ when the condition for the thermal-timescale MT (TTMT) is no longer satisfied.
For instance, we find that if a $1.2\,M_\odot$ giant with a $0.37\,M_\odot$ core  has a companion of $1.1\,M_\odot$ and MT proceeds with $\beta=0.3$, 
then  $\zeta_\mathrm{RL}= \zeta_\mathrm{eq} $  when the mass ratio in the system becomes 0.75 (i.e., when the donor mass is decreased to $\sim 0.9\,M_\odot$).
Due to some inertia in the star's response, the TTMT may proceed afterwards,  but no longer than for $\sim \tau_\mathrm{TH}$, and at 
$q_\mathrm{crit}\approx 0.75$ the MT rate is likely to be at its maximum. In the case of a $1\,M_\odot$ companion, the maximum TTMT rate 
is expected to occur at $q_\mathrm{crit}\approx 0.85$ (donor mass is decreased to $1\,M_\odot$).
Though it does not yet give us strong predictive power on when the TTMT must stop,
it helps us to understand the results of our simulations qualitatively. 
\section{The First Phase of Mass Loss}
\label{1stMassLoss}
\subsection{Stellar-evolution code}

In order to evaluate our model we perform numerical calculations using the \texttt{ev}\footnote{The current 
version of \texttt{ev} is obtainable on request from \texttt{eggleton1@llnl.gov}, along with data files and a 
user manual.} binary stellar-evolution code originally developed by Eggleton 
\citep[][and references therein]{Eggleton1971, Eggleton1972, YakutEgg2005} and updated 
as described in \citet{PTEH1995} and \citet{GPH2008}. The code solves the 
equations of stellar structure and evolution for both components of a binary simultaneously using an implicit 
scheme over an adaptive mesh. We model each star using a grid size of 200 mesh points, as this consistently 
provides stable results. This results in the model being quite stable against very short-timescale 
instabilities, allowing us to quickly evolve our models up the asymptotic giant branch (AGB). Simultaneous 
calculation of both components is essential in order to account for non-conservative mass transfer, as each 
star's evolution is no longer separable from the other \citep{EKE2002,YakutEgg2005}.
	
Opacity tables are taken from OPAL \citep{IRW1992}, with the low-temperature range taken from \citet{AF1994}. 
A diffusion equation models convective mixing for each of the composition variables, and 
overshooting is modelled as in \citet{SPE1997}. The helium flash in the degenerate core of low-mass stars 
is avoided by substituting the stellar model with one in which helium has just been ignited in the core. 
The initial metallicity for all models is assumed to be roughly solar ($X = 0.70, Y = 0.28, Z = 0.02$).

We account for both RLOF and a stellar wind in the boundary conditions for the mass, assuming a Reimers-like 
\citep{Reimers1975, Reimers} model for stellar-wind-driven mass loss: 
\begin{equation}
  \dot{M} = 4.0\times10^{-13} \, \eta \, \frac{R}{R_\odot}\frac{L}{L_\odot}\frac{M_\odot}{M} \, M_\odot \mathrm{yr}^{-1},
  \label{eq:wind}
\end{equation}
where we set $\eta = 0.2$.  

\begin{figure}
  \includegraphics[height=.25\textheight]{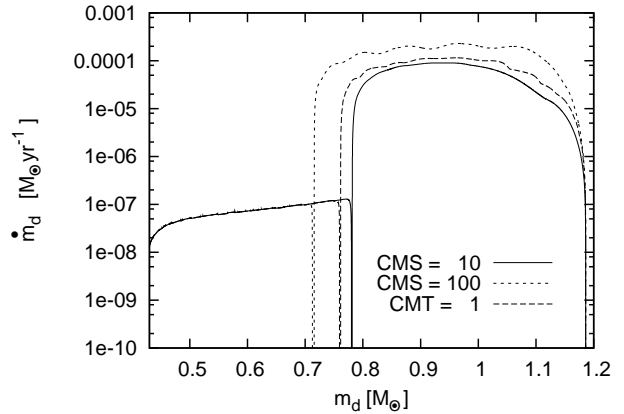} 
 \caption{Mass-transfer rates according to different prescriptions, for a $1.2+1.1\,M_\odot$ RG-MS binary, with a period at the start of RLOF of 100\,d.
   The solid and dotted lines indicate the CMS formulation, the dashed line is for the CMT formulation.  A mesh of 200 points is used in each case.  
   \label{fig:comp_mesh} 
 } 
\end{figure}

Eggleton's code provides two means of parametrizing the mass loss from the donor during RLOF. 
One way is to allow the donor's radius to slightly exceed its Roche lobe, whereupon the mass loss is 
computed in terms of this excess:
\begin{equation}
  \dot M_\mathrm{mt} = \mathrm{CMS} \times \left[\log \left(\frac{r}{r_\mathrm{RL}}\right)\right]^3
\end{equation}
Here CMS is an arbitrary parameter; a larger value is expected to provide more appropriate 
mass-transfer rates, as this allows for a smaller Roche-lobe overflow throughout the mass-transfer evolution, and hence smaller time steps taken by the code.
A lower value may provide MT rates that are too small for the case of TTMT, 
though the donor is exceeding its Roche lobe more than in the case of a bigger value, and the code takes larger time steps.
In Fig.\,\ref{fig:comp_mesh}, we plot the evolution of the mass-transfer rate for 
a $1.2+1.1\,M_\odot$ binary with an initial period of 100 days while varying CMS. 
We find that so long as a reasonable value is chosen (i.e. one for which runaway MT is not induced artificially), 
the same salient evolutionary features are seen, albeit at slightly different moments in the MT history. 
Notably, in all cases MT enters a final phase of slow, steady nuclear-timescale MT following 
a thermal-timescale phase and a pause (see below). 
In addition, for each case final values of $m_\mathrm{d,c}$ and P agree within $\sim 1\%$. 
We adopt CMS = 10 as our default value --- it is both found to be the most appropriate value in our calculations, 
and is generally recommended for donor stars in the mass range considered here (Eggleton, private communication).

An alternative is a brute-force formulation, 
labelled in the code as CMT, in which the potential is calculated 
at each grid point within the donor star. 
This is then used to calculate mass flux $\xi$ away from the donor 
as a function of depth, zero below the $L_1$ surface (again, 
the donor radius is allowed to slightly exceed its Roche lobe). 
The MT rate is then calculated as $\dot M = \mathrm{CMT} \cdot \xi$, 
where CMT is an arbitrary parameter. This method is designed for contact 
binaries. 
In Fig.\,\ref{fig:comp_mesh}, we plot the MT evolution for the same system using this prescription: for CMT = 1, 
we find a MT rate history comparable with that obtained for CMS = 10, but with more rapid oscillations. 
At higher values we find runaway MT which we again assume to be artificial, as they seem to be the result of the arbitrary coefficient.

The oscillatory pulsations noted above, when a result of numerical error, 
arise as an artefact of defining too large mass shells when implementing 
the mesh used to compute our model, partly due to our choice of a relatively small number of mesh points (200). 
This causes the boundary between the star's convective envelope and the core to vary rapidly at high MT rates, 
resulting from overly large mass shells changing back and forth between a convective 
and a non-convective state at the core boundary. These pulsations appear to be much 
stronger when using the CMT prescription (Fig.\,\ref{fig:comp_mesh}), 
suggesting greater instability. However, these pulsations do not have a 
strong influence on the evolution of our model and its outcome, and for our purposes may be ignored.

\subsection{A maximum conservation factor}    

Whether the dynamical-stability condition $\zeta_\mathrm{ad} > \zeta_\mathrm{RL}$ is satisfied or not depends on $\beta$: 
$\zeta_\mathrm{RL}$ is dependent on $\beta$ through the responses of both the mass ratio and the orbital separation
to MT (see Eqs.\,\ref{eq:zeta_rl}, \ref{eq:dlna_dlnm} and \ref{eq:dlnq_dlnm}).
As $\zeta_\mathrm{ad}$ and $\zeta_\mathrm{RL}$ are both monotonic functions of mass loss,  it is a sufficient condition 
that MT be stable at the onset of RLOF in order to ensure stability throughout. 
We reject system configurations in which this condition is violated. 

We realise that there is a difference between a true dynamical stability (or instability) --- the one arises
on time-scales shorter than dynamical-timescale, before a star obtain hydrostatic equilibrium ---
and the dynamical stability used commonly for the purpose of studies of mass transfer stability.
The latter one, as described earlier in \S4, operates on timescales longer than the dynamical timescale
(as all the analyzed stellar model are in hydrodynamical equilibrium) but much shorter than thermal 
timescale, so the entropy of the stellar layers is not changed.
The code we use  always generating stellar models in hydrostatic equilibrium,
and as such is not capable of modelling a true dynamical-timescale instability,
however  is able work on a timescale much shorter than is required to change the entropy.
As such, through out mass transfer sequences, when the rate of the MT exceeds TTMT, 
we efficiently obtain  $\zeta_\mathrm{ad}$ as discussed above, but for actual stellar models 
(instead of composite polytropes).
We find that with $\beta$ close to the maximum stable value, 
we obtain MT rates exceeding the thermal-timescale MT rate, 
although this is achieved only after $\sim \tau_\mathrm{th}$ has passed after the onset of RLOF.  

\begin{figure}
  \includegraphics[height=.25\textheight]{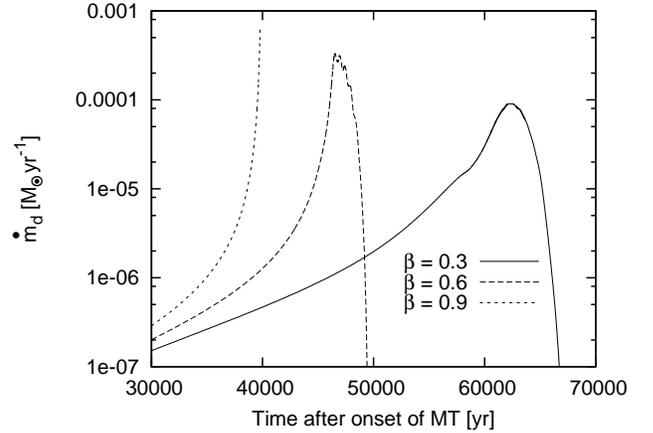}
  \caption{
    Mass-transfer rates for different conservation factors $\beta$ in a binary of 
    $m_\mathrm{d,i} = 1.2\,M_\odot$, $m_\mathrm{a,i} = 1.1\,M_\odot$, and with an orbital period 
    at the beginning of RLOF of 100\,d. 
    \label{fig:comp_MT}}
\end{figure}

As an example, we model the case of a $1.2\,M_\odot$ red giant with a $1.1\,M_\odot$ 
companion, with an initial 
period of 100 days, which initiates MT at a core mass of approximately $0.345\,M_\odot$ and with a donor radius of $46.4\,R_\odot$. 
For a conservation factor greater than $\beta_\mathrm{max} \approx 0.89$, 
runaway MT ensues (see Fig.\,\ref{fig:comp_MT}):
MT stability is unable to recover on the thermal timescale 
for $\beta = 0.9$, leading to the code crashing. 
We interpret this as indicative of a runaway MT event. 
In this case the MT rate approaches and finally exceeds the thermal-timescale-MT rate. 
Note that the maximum conservation factor found here is considerably higher than that found 
when taking the donor as a condensed polytrope considered with the core and the total masses as in the described above giant, in which case $\beta_\mathrm{max} \approx 0.32$. 

\subsection{Detailed evolution} 

Note from Fig.\,\ref{fig:comp_mesh} that the MT phase appears to pass through two distinct sub-stages. 
Using the example above (a $1.2+1.1\,M_\odot$ progenitor system with $P_\mathrm{i}=100$\,d) 
for the case $\beta = 0.3$, we observe that there are in fact three such stages. 

\begin{figure}
  \includegraphics[height=.25\textheight]{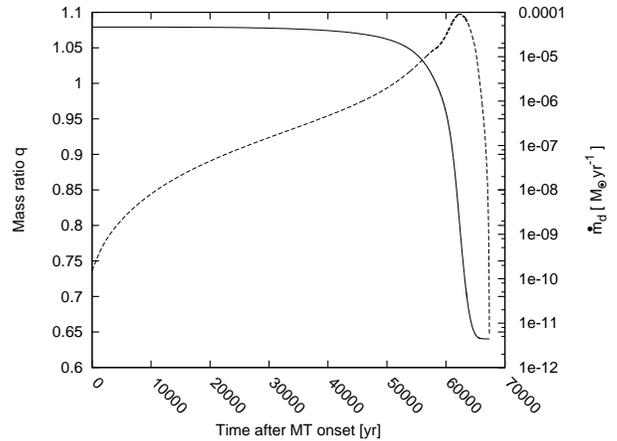}
  \caption{
    Time evolution of the MT rate (dashed line) and mass ratio (solid line) after the onset of MT.
    The MT proceeds on the thermal timescale, and $q_\mathrm{crit} \approx 0.75$.}
  \label{fig:1stphaseend} 
\end{figure} 

At the onset, the MT rapidly accelerates to a timescale comparable to the donor's thermal timescale, 
reaching a dramatic peak $\dot M$ before rapidly falling off. This turn-off occurs approximately 
$\tau_\mathrm{th}$ after the condition $\zeta_\mathrm{RL} (q) < \zeta_\mathrm{eq}$ is met (see 
Fig.\,\ref{fig:1stphaseend}), as the mass ratio reaches $q_\mathrm{crit}$. This initial stage strips 
the envelope down (removing $\sim 50-70\%$ of the envelope mass), to which the donor reacts by 
contracting until the binary becomes detached (underfilling the Roche lobe by $\sim 5 \%$). 
This results in a pause in MT lasting for $\sim 1-4$\,Myr (this recovery time is approximately the 
thermal time of the whole donor, whereas the thermal time of 
only its outers layers played a role in the determination of the TTMT rate), during which time 
the donor's core grows slightly until RLOF resumes.
 
This pause is followed by another stage of MT on the star's nuclear timescale, in which the remainder 
of the envelope is transferred, save for a final $\sim 10^{-3}\,M_\odot$.
While often neglected \citep{Han1998}, in the 
course of the pause and nuclear-timescale phase of MT the donor's core mass can increase by $\sim 20-30\%$, having a profound effect on the outcome of the system.

As for the donor's remaining envelope mass, at the end of MT this collapses (on its thermal timescale) 
onto the surface of the degenerate core \citep{DS70,Ivanova10, Justham10}. The same behavior is observed 
in our model, and the amount that remains in the envelope at the end of MT is generally $\sim 3 \%$ of 
the original envelope mass. This means that for smaller core masses, we see thicker envelopes remaining. However, using our code it is difficult to make a precise estimate of the final mass of the 
envelope remnant, as our model breaks down at the collapse of the envelope for low core masses. 
For the purpose of this study, we assume that the MT ends very shortly after the code breaks down, and that the 
remaining mass in the envelope will be burned, i.e., we take the total mass at the end of the MT phase as our final 
WD mass. We note that this may introduce a slight bias in the results, pushing the final mass ratio somewhat 
closer to unity and the final period to very slightly shorter values in the least massive systems. 

\begin{figure}
  \includegraphics[height=.25\textheight]{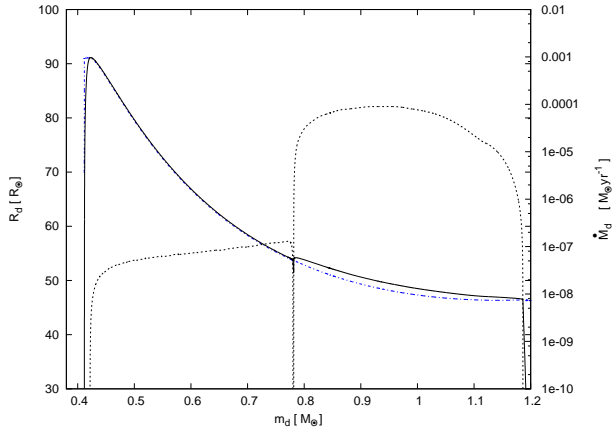}
  \caption{Evolution of the donor radius (solid line) and MT rate (dotted line) as a function of donor mass during MT, 
    for an initially $1.2+1.1\,M_\odot$ binary with an initial period of 100 days. The Roche-lobe radius is denoted 
    by dashed blue line.}
  \label{rad_ev}
\end{figure}

\begin{figure}
  \includegraphics[height=.25\textheight]{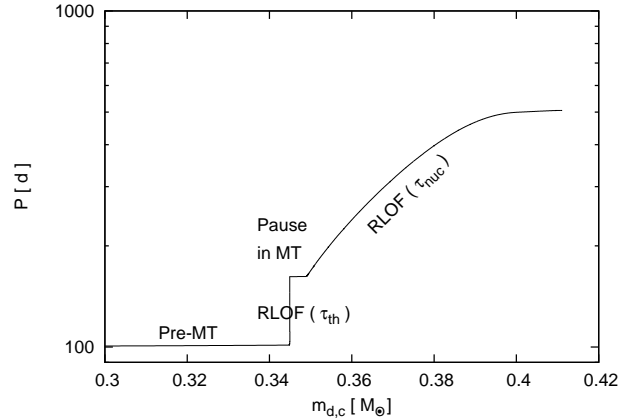}
  \caption{Period growth with core-mass evolution for a $1.2+1.1\,M_\odot$ binary with an initial period of 100\,d, 
    broken down in the three stages of the first MT phase.} 
  \label{mdc_P}
\end{figure}

\begin{figure}
  \includegraphics[height=.25\textheight]{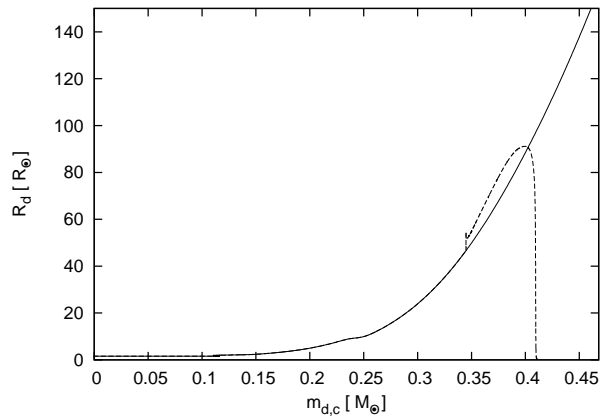}
  \caption{Deviation from the core mass--radius relation for the donor star from the system in Fig.\,\ref{rad_ev}. 
    Solid line: unperturbed star, dashed line: perturbed donor star.}
  \label{rad_diff}
\end{figure}

In order to better illustrate each stage of the first MT phase, we display the evolution of the MT rate and 
donor radius as a function of the donor mass in Fig.\,\ref{rad_ev}, and the evolution of the orbital period as a function 
of the donor's core mass in Fig.\,\ref{mdc_P}. Figure~\ref{rad_diff} shows the donor's radius as a function of 
its core mass, and we note that the radius of the donor during the nuclear-timescale MT is significantly larger 
than what we would expect of an unperturbed star. 

\begin{figure}
  \includegraphics[height=.25\textheight]{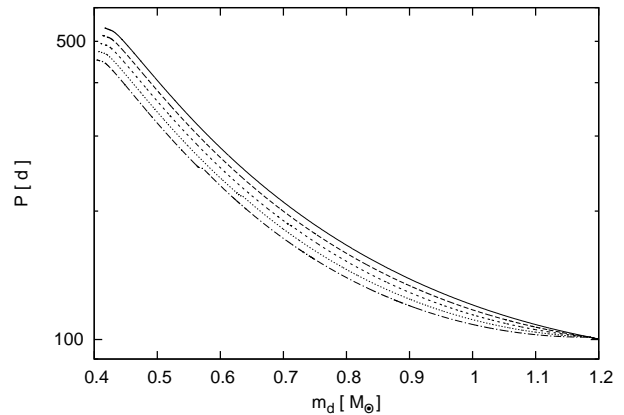}
  \caption{Evolutionary tracks for a $1.2+1.1\,M_\odot$ binary with $P_\mathrm{i} = 100$\,d through stable, 
    non-conservative MT, for $\beta=$ 0 (top), 0.2, 0.4, 0.6, and 0.8 (bottom).
    \label{fig:endstate:beta}}
\end{figure}

\begin{figure}
  \includegraphics[height=.25\textheight]{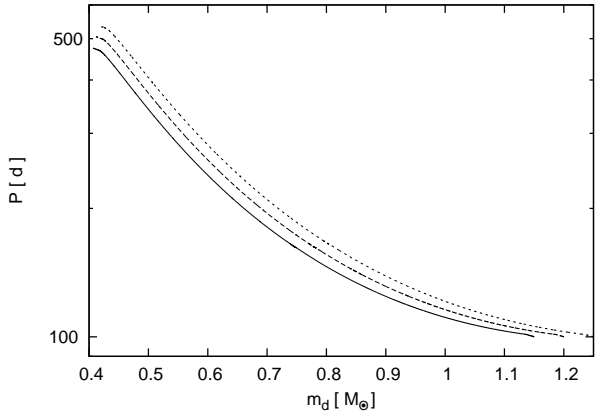}
  \caption{Evolutionary tracks for an initial binary system with $P_\mathrm{i} = 100$\,d through stable, non-conservative MT, 
    for initial $m_\mathrm{d} = 1.15\,M_\odot$ (bottom), $1.20\,M_\odot$, $1.25\,M_\odot$ (top), and a constant mass ratio $q \approx 1.091$.
    \label{fig:endstate:mdi}}
\end{figure}

\begin{figure}
  \includegraphics[height=.25\textheight]{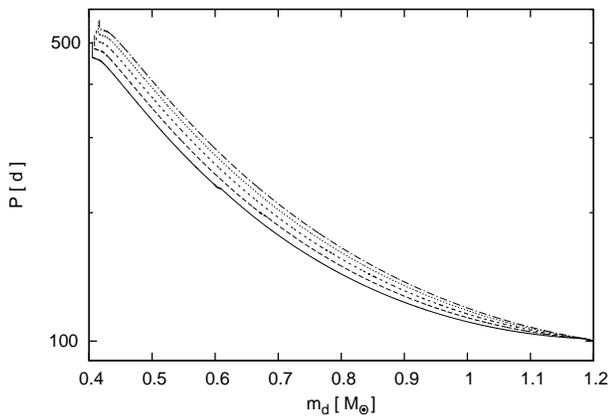}
  \caption{ Evolutionary tracks for an initial binary system with $P_\mathrm{i} = 100$\,d through stable, non-conservative MT, 
    for $q_\mathrm{i} =  1.200$ (bottom), 1.143, 1.091, 1.043, and 1.008 (top). 
    \label{fig:endstate:qi}}
\end{figure}

To investigate how the above evolution varies under different initial parameters, we now recompute the evolution of
our $1.2+1.1\,M_\odot$ binary with $P_\mathrm{i} = 100$\,d, and vary the conservation factor (Fig.\,\ref{fig:endstate:beta}), 
initial donor mass (Fig.\,\ref{fig:endstate:mdi}), and initial mass ratio (Fig.\,\ref{fig:endstate:qi}). This exercise
demonstrates that for initial conditions leading to the first phase of MT,
\begin{itemize}
\item the conservation factor is anti-correlated with the final period,
\item a larger initial donor mass (for constant initial mass ratio) leads to a longer final period,
\item the initial mass ratio is anti-correlated with the final period.
\end{itemize}

In the next section we will also vary the initial period, which will naturally lead to larger core masses for larger 
initial separations. In addition, we will compute the further evolution of such binaries through the 
RLOF of the (former) secondary, which --- due to the dramatic change in mass ratio --- will proceed in an unstable manner, 
leading to a typical CE.

\section{The Second Phase of Mass Loss}

\subsection{An Ensuing Common Envelope}
In the intervening period following the end of the first phase of MT, further orbital expansion may be driven by a wind 
from the (former) secondary (i.e., the originally less massive component).
However, as noted above, in many cases our code has difficulty modelling the collapse of the 
envelope at the end of stable MT, and therefore cannot progress through this intermediate stage. Hence, in continuing 
to the second phase of mass loss we ignore wind loss from the secondary. Those models which do survive the end of the 
initial MT phase suggest the relative error introduced in the intermediate period ($P_\mathrm{m}$) by ignoring the wind-driven expansion 
is on the order of $5-10\%$. For the subsequent CE phase, the resulting error in the radius at RLOF is small, as at the onset of 
RLOF the radius will only be related to the period weakly through $P^2\sim a^3 \sim R^3$. Since inverting the $m_\mathrm{c}$--$R$ 
relation gives the core mass a roughly logarithmic dependence on the radius, we may take the error in the final core mass of the 
secondary at the onset of the CE as being negligible. As the post-CE separation is directly proportional to the initial separation 
in the $\alpha_\mathrm{CE}$ formalism, we may take the relative error in $P_\mathrm{Post-CE}$ as being roughly the same as that 
in $P_\mathrm{m}$. 

From this point we can easily estimate the resulting final binary configuration for a given system after a second episode 
of mass loss in which a common-envelope phase ensues (assuming that the $\alpha_\mathrm{CE}$ formalism holds, and computing 
the orbital change directly from the binding energy in our models). We use a core-mass radius relation from the parametrized 
stellar models of \citep{Hurley_mcr} in order to estimate the core mass of the secondary at which the onset of a CE takes place, 
as this approximates the radii of all our models to within a few percent. As a CE event is a dynamical-timescale process, we 
neglect core-mass evolution over the course of the CE phase.  

\begin{figure}
  \includegraphics[height=.25\textheight]{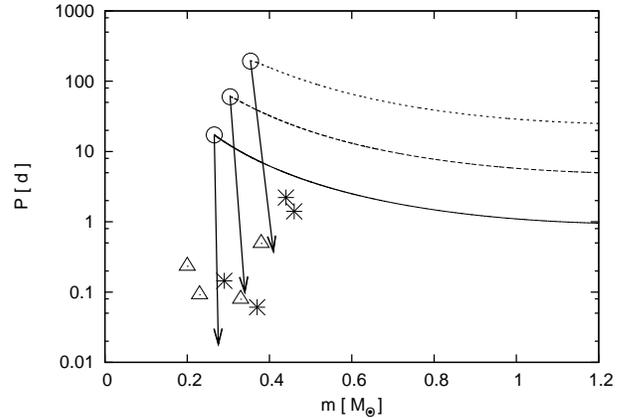}
  \caption{Full evolutionary tracks for a binary with initial masses of $1.2+1.1\,M_\odot$ through an initial stage of stable MT followed by a CE, 
    for $P_\mathrm{i} = $ 1, 5, and 25\,d with respect to donor mass. Circles indicate remaining (core) mass of primary and orbital 
    period after the system has gone through the initial phase of stable MT.
    Arrow endpoints indicate the final (WD) mass of the secondary and orbital period after the second, unstable episode of mass loss (CE phase). 
    Hence, an arrow slanted to the right indicates a final system where the secondary WD is more massive than the primary.
    Black lines trace the evolution of the system through the first (dashed) and second (solid) phases of mass transfer. 
    Black stars (from double-lined binaries, table 1) and triangles (from \cite{Kilic2010}) mark parameters of observed DWDs 
    (orbital period and mass of the inferred secondary, see Tables~\ref{table:WD}, \ref{table:WD2}).
    \label{fig:endstate:period}}
\end{figure}

\begin{figure}
  \includegraphics[height=.25\textheight]{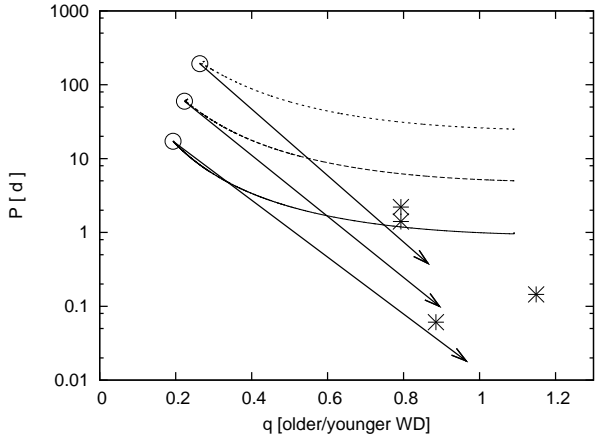}
  \caption{ Full evolutionary tracks for a binary with initial masses of $1.2+1.1\,M_\odot$ through an initial stage of stable MT followed by a CE, 
    for $P_\mathrm{i} = $ 1, 5, and 25\,d, plotted against mass ratio.  Lines and symbols are as in Fig.\,\ref{fig:endstate:period}. Due to the much 
    greater uncertainty in the masses of the companion WDs in Table~\ref{table:WD2}, we plot only those observed systems 
    listed in Table~\ref{table:WD}.
    \label{fig:endstate:q_P}}
\end{figure}

As an example, we evolve a set of $1.2+1.1\,M_\odot$ systems with $\beta = 0.3$ and varying initial period as above through 
a second, unstable episode of MT, plotted in Figs.\,\ref{fig:endstate:period}, \ref{fig:endstate:q_P}. These form a fairly 
linear track of end products in both $m-\log P$ space and $q-\log P$ space, with the final periods, mass ratios, and core 
masses of the inferred younger companions matching nicely with observations. Initial period aside, the final period is most 
strongly dependent on the efficiency of the CE, the lower values of $\alpha_\mathrm{CE}$ driving the final separation to 
smaller values. Further modulation of these results may come from varying the initial donor mass, initial mass ratio, and 
conservation factor.

\subsection{An Exception: WD 1101+364}

Among those double-lined DWDs in which neither component's progenitor has undergone the He flash, we determine then that the 
stable MT+CE channel provides a reasonable explanation for the formation of those systems with q( inferred older/younger) $<$ 1. 
This excludes only one system, WD\,1101+364 (Fig.\,\ref{fig:endstate:q_P},) whose mass ratio is greater than unity, implying the 
orbit contracted with the loss of the initial donor's envelope. Attempts to model this 
system as the product of a double-CE \citep{Nelemans00, Sluys2006} found this could only be done with an unphysical value for 
$\alpha \lambda$ for the first mass loss phase; this was part of the motivation for the development of the $\gamma$-formalism. 
\cite{Nelemans00} suggested this system could better be modelled as the product of two $\gamma$-prescribed events, with the 
initial mass inferred from this reconstruction implying a difference in cooling ages of $\sim 800$\,Myr.
However, \cite{Maxted02_2} found an observed difference of $\sim 215$\,Myr, certainly irreconcilable with the predicted value. 
As they are quick to point out though, cooling models for helium white dwarfs remain quite uncertain, 
and drawing strong conclusions regarding their relative ages remains a difficult task. 

Compounding the problem of determining WD ages is the prospect of rejuvenation, as even the accretion of interstellar matter may 
pose a significant problem \citep{rejuvenate}. It is therefore 
quite possible that even our understanding of which component of any DWD is older may be incorrect. Indeed, WD ages are based 
on models of cooling tracks whose uncertainties are highly dependent on the size of the remaining hydrogen envelope, with 
thicker envelopes working as `blankets' and causing WDs to appear hotter, hence younger. WD\,1101+364 has the smallest component 
of any of the companion masses, and 
as we have seen above ($\S5$) this corresponds to a thicker anticipated envelope. Such a turnaround in the respective age difference 
is not impossible. We note that a reversal of the mass ratio would place WD\,1101+364 almost directly on 
the line traced by our example track, and \cite{Han1998} models this system as originating through a stable MT+CE scenario under 
similar assumptions, though this is hardly definitive. As well, we have neglected the effect of hydrogen-shell flashes after the 
collapse of the envelope here 
\citep[e.g.\ Sect.\,5.3 in][]{Yoon04}, which may reduce the envelope mass significantly due to nuclear burning, mass 
transfer and mass loss.  Shell flashes occur earlier and are stronger for lower-mass proto-WDs.

Further modelling of WD 1101+364 in \citet{Sluys2006} produced better results 
by varying the means by which the binary shed angular momentum. Assuming that in the first phase mass is lost directly from 
the donor in an isotropic wind, while in the second phase it is re-emitted from the accretor, gave a small set of solutions 
which were nearly within error. Yet a physical explanation for such processes to occur on a 
timescale short compared to the nuclear-evolution timescale, as assumed,\footnote{Note that while \citet{Sluys2006} \emph{claim} to assume
that envelope ejection described by the $\gamma$-prescription must occur on the dynamical timescale, they in fact make a looser assumption; for the mechanism to work, ejection of the envelope
on a timescale shorter than the nuclear-evolution timescale --- so that the core mass doesn't change --- is sufficient, and this is their \emph{actual} assumption.} 
remains as of yet elusive. At best, we can say that something appears to be amiss in our 
understanding of WD\,1101+364.

\section{Discussion and Conclusions}

In this paper we addressed the problem of double-white-dwarf formation,
beginning with an analysis of the applicability of the previously proposed $\gamma$-formalism  for an arbitrary binary.
Originally, the $\gamma$-formalism was developed in order to describe a dynamical-timescale phase of mass loss in terms 
of a {\it single}-valued parameter for all possible binaries, without a specific
assumption of the physical characteristics of the evolution. This
value was then found based on the study of currently observed DWD systems.

We have demonstrated that, when parametrizing the consequent angular momentum loss with a single value ($\gamma$), 
two dynamical-timescale phases of mass loss can only model the formation of arbitrary DWD binaries with the use of 
{\it different} values of $\gamma$ which are {\it very finely tuned} for every mass ratio. 
The underlying physics of a first phase of dynamical-timescale mass loss in such a process,  
in which the envelope of the initial primary must be expelled without significant 
shrinkage (or even widening) of the orbit, remains without an explanation under the formalism of a $\gamma$-prescribed envelope removal. 
However, the apparent violation of energy conservation during the 
first phase of mass loss as described by the $\gamma$-formalism in many cases indicates that the envelope ejection cannot be driven by any dynamical
process in such instances, but instead appears to be mimicking the stable, non-conservative MT that runs on a longer timescale ---
thermal or nuclear. 

Indeed, in $\S 5,6$ we demonstrate that such stable, non-conservative MT provides a physically 
motivated, easily implemented mechanism in lieu of any parametrization of the energetics or angular momentum balance 
during the first mass loss phase, which here obscures the physical interpretation 
and, as we have seen, requires careful fine tuning. Circumventing the need for any single-valued parametrization, 
we compute the evolution of any arbitrary low-mass DWD progenitor for which the inferred older 
component has not undergone the helium flash through the self-consistent 
computation of the evolution of a binary through stable MT (assuming the stability conditions outlined in $\S4,5$ hold), followed then by a CE.

Hence, this scenario allows us to reproduce the period and mass ratio of the observed DWDs with low-mass companions 
($M \lesssim 0.46\,M_\odot$) with only a single combination of initial binary-component masses ($1.2+1.1\,M_\odot$).
This is very promising, but more work is needed to determine whether we can indeed explain all the observed low-mass 
DWDs in which the older component is less massive, 
including the differences in cooling age using this scenario. The latter, of course, would require stronger constraints 
on WD cooling tracks (including the effect of rejuvenation) than exists now.

For MT evolution, we analyze how non-conservative 
MT should be in order to proceed stably.
We did not discuss in detail the exact mechanism for expelling mass from the system during MT,
and do not propose what the exact value of the efficiency should be, but rather study the consequences
of non-conservation during MT with an adopted conservation factor.
We may argue however that MT is rather likely to be non-conservative, especially during the TTMT phase.
The strongest argument for this is due to the very high MT rate expected: some of the binary systems, at the peak of TTMT, 
could have MT rates above the Eddington limit ($\sim 10^{-3}\,M_\odot \mathrm{yr}^{-1}$ for MS accretors 
in the mass range we considered).
Observationally, though, it is hard to make any constraint: the TTMT phase is of course very short lived.
We note however that among observed binaries undergoing MT with giant donors, e.g., in  symbiotic stars,  
there are observed systems showing powerful jets --- clearly, the MT is not fully conservative there (e.g., 
in symbiotic stars like CI Cygni \citep{CICYG}, CH Cygni \citep{CHCyg}), even though the MT proceeds there on a nuclear timescale.
Young pre-MS stars are the only observed non-compact objects that accrete at rates almost 
comparable to our TTMT rates, and are known to have jets as well e.g., HH 30 \citep{HH30}. 
Thus our assumption that bipolar re-emission from the accretor drives non-conservation in MT, and therefore the specific 
angular momentum of the lost material is that of the accretor, seems at least in principle quite reasonable.

In the context of earlier studies, it is important to note that the 
progenitor donors with $M \gtrsim 2\,M_\odot$ often considered 
previously would not be able to form a WD immediately following an episode of MT --- they would instead form a low-mass He 
star with a lifetime of up to 0.5\,Gyr \citep{Yungelson08}, relevant to the age difference of many of the solutions described in table 5 of 
\cite{Sluys2006}. In this study we find that in the low-mass regime the stable MT + CE channel, with an initially 
$1.0-1.3\,M_\odot$ donor, provides a natural explanation for all but one of the observed double-white-dwarf binaries for 
which the inferred older component is below $0.46\,M_\odot$.

Thus, systems such as PG\,1115+116, in which the younger companion is the more massive, can easily form as a result of the orbital 
expansion during the initial phase of RLOF, followed by a CE. The relatively small delay between the two phases of mass loss 
from the system, inferred from observations, is also easily justified by the accelerated evolution of the accretor as a result 
of the first MT episode. Of course, a strong determination of the relative ages in a DWD is a difficult task, further confounded 
by the possibility of rejuvenation. 

The system WD\,1101+364 presents a clear challenge in that its older component is the more massive, yet it cannot be explained as 
the product of a double common envelope without resorting to unphysical efficiencies in the first phase of mass loss. One possibility 
may be some intermediary process, such as that suggested by \cite{Nelemans00}, in which the envelope of the initial primary is lost 
without significant orbital shrinkage; \cite{Sluys2006} suggests this stage may be understood as a phase of very rapid wind loss. 
However, an obvious mechanism for this remains lacking. It is also possible that the formation 
of a small circumbinary disk in the first phase of mass transfer may allow for some shrinkage of the orbit, however there is a lack 
of evidence at the present moment to support such a hypothesis. Perhaps the most likely explanation, given the small secondary mass 
in WD\,1101+364 and therefore its thicker hydrogen envelope, is that the resulting uncertainty in the cooling tracks means that the 
$0.29\,M_\odot$ companion is older than previously thought.

\acknowledgements{
  We thank P.P.\ Eggleton and E.\ Glebbeek for making their binary-evolution code available to us, as well as Gijs Nelemans and Craig Heinke for helpful discussion.  
  NI acknowledges support from NSERC and Canada Research Chairs Program.
  MvdS acknowledges support from a CITA National Fellowship to the University of Alberta, and support from the Dutch Foundation for Fundamental Research on Matter. 
}
\bibliography{comenv}
\bibliographystyle{apj}

\begin{table}[ht]
\caption{Double White Dwarfs with Known Masses and Periods}
\begin{center}
\begin{tabular}{c c c c c c}
\hline\hline

& System & $M_1 [M_\odot]$ & $M_2 [M_\odot]$ & Period [d] & q\\ [0.5ex]
\hline
1 & WD\,0136+768 & 0.35 $\pm$ 0.02 & 0.46 $\pm$ 0.03 & 1.407227 & 0.793\\
2 & WD\,0957-666 & 0.32 $\pm$ 0.02 & 0.37 $\pm$ 0.02 & 0.06099 & 0.885\\

3 & WD\,1349+144 & 0.35 & 0.44 & 2.2094 & 0.793\\
3 & WD\,1349+144*& 0.44 & 0.44 & 2.2094 & 0.793*\\

4 & WD\,1101+364 & 0.35 & 0.29 & 0.14458 & 1.149\\
\hline
\end{tabular}
\end{center}

\tablecomments{Known double-lined DWD systems with established orbital parameters, for which the (inferred) older WD ($M_1$) 
  is less than $0.46\,M_\odot$. Results taken from \cite{Bragaglia09} [2], \cite{Marsh95} [4], \cite{Moran97} [2], \cite{Maxted02_2} [1,2], 
  \cite{SPY03} [3]. For WD 1349+144, we note that the measured mass of $M_1$ is $0.44\,M_\odot$ (marked with an asterisk), however based on 
  the (much more trusted) observed q-value of $q \approx 0.793$, $M_1 \approx 0.35$. We plot the latter. Also note that here we define 
  $q$ = (inferred older WD)/(inferred younger WD), opposite to the usual convention. \label{table:WD}}
\end{table}

\begin{table}[ht]
\caption{Double White Dwarfs with Partially Known Masses and Periods}
\begin{center}
\begin{tabular}{c c c c c}
\hline\hline
System & $M_{1,\mathrm{min}} [M_\odot]$ & $M_{1,\mathrm{max}} [M_\odot]$ & $M_2 [M_\odot]$ & Period [d]\\ [0.5ex]
J0022+0031 & 0.35 &   ?  & 0.38 & 0.492 \\
J0022-1014 & 0.19 & 0.45 & 0.33 & 0.079 \\
J1234-0228 & 0.09 & 0.22 & 0.23 & 0.092 \\
J1625+3632 & 0.08 & 0.17 & 0.20 & 0.233 \\
\hline

\end{tabular}
\end{center}
\tablecomments{Known single-lined DWD systems with established period and mass of one component. Where possible, 
  upper and lower bounds on the companion masses are taken from the source paper (to varying confidence levels). 
  Results taken from \cite{Kilic2010}. The observed white dwarf is presumed to be the younger one ($M_2$).
  \label{table:WD2}}
\end{table}

\end{document}